\documentclass[11pt,twoside]{gsajnl}

\articletype{inv} 

\runningtitle{BBChain | Computação confidencial} 
\runningauthor{de Avellar \textit{et al.}}

\title{A visão da BBChain sobre o contexto tecnológico subjacente à adoção do Real Digital}

\author[1,$\ast$]{M. G. B. de Avellar}
\author[1$\bullet$]{A. A. S. Junior}
\author[1$\circ$]{A. H. G. Lopes}
\author[1$\diamond$]{A. L. S. Carneiro}
\author[2$\star$]{J. A. Pereira}
\author[2$\ddagger$]{D. C. B. D. da Cunha}

\affil[1]{BBChain | Avenida Juscelino Kubitschek, 1327 - 4º andar – Conjunto 41 CV 1152 – Vila Nova Conceição CEP: 04543-011 – São Paulo, SP}
\affil[2]{Microsoft Corporation | Avenida Juscelino Kubitschek, 1909 - 16º andar - Vila Nova Conceição CEP: 04543-900 - São Paulo, SP}

\correspondingauthoraffiliation[$\ast$]{~marcio.g.avellar@bbchain.com.br. \\ 
$\bullet$~alexandre.junior@bbchain.com.br \\
$\circ$~andre.lopes@bbchain.com.br \\
$\diamond$~carneiro@bbchain.com.br \\
$\star$~jopereira@microsoft.com \\
$\ddagger$~davicunha@microsoft.com
}

\begin{abstract}
Neste artigo, exploramos o uso da computação confidencial no contexto das CBDCs, utilizando o framework CCF da Microsoft como exemplo. Através do desenvolvimento de uma experimentos e da comparação de diferentes abordagens e métricas de desempenho e segurança, buscamos avaliar a eficácia da computação confidencial para melhorar a privacidade, segurança e desempenho das CBDCs. Os resultados preliminares de nossa pesquisa sugerem que a computação confidencial pode ser uma solução promissora para os desafios tecnológicos enfrentados pelas CBDCs. Ao implementar a computação confidencial em DLTs, como o Hyperledger Besu, e utilizar frameworks como o CCF, podemos aumentar a confidencialidade e a privacidade das transações, ao mesmo tempo em que mantemos a escalabilidade e a interoperabilidade necessárias para um sistema financeiro global e digital. Em conclusão, a computação confidencial tem o potencial de desempenhar um papel fundamental no desenvolvimento e implantação de CBDCs, contribuindo para um futuro mais seguro, privado e eficiente no mercado financeiro. 
\end{abstract}


\dates{\rec{xx xx, xxxx} \acc{xx xx, xxxx}}

\begin{document}

\maketitle

\thispagestyle{firststyle}
\vspace{-13pt}

\lettrine[lines=2]{\color{color2}E}{}sse artigo se encontra dentro do subeixo temático {\it Tecnologia da Informação} e foi submetido para a {\it Chamada para Submissão de Artigos | A tokenização das finanças: dos criptoativos às moedas digitais de bancos centrais}.

\section{Introdução}

O Banco Central do Brasil iniciou, em 06.03.2023, a fase de pilotos do Real Digital, a versão tokenizada do Real, a moeda fiduciária brasileira {\bf \textcolor{myGreen03}{\citep{sciarretta_01,bcb_03}}}. A previsão é de que o Real Digital esteja disponível para determinados segmentos até fim de 2024. O Real Digital é uma CBDC ({\it Central Bank Digital Currency}), uma moeda digital emitida pelo Banco Central de um país e cujo propósito é ser a forma digital da moeda nacional, sendo, portanto, regulada pelo governo. CBDCs diferem de {\it stablecoins}, que têm valor ligado a um ativo referencial e são emitidas por instituições financeiras e também de {\it criptocurrencies}, como a Bitcoin {\bf \textcolor{myGreen03}{\citep{nakamoto2009bitcoin}}} e o Ethereum {\bf \textcolor{myGreen03}{\citep{ethereumWhitepaper}}}, emitidas por qualquer entidade e cujo valor é flutuante, variando segundo as forças do mercado.

Essa iniciativa vem na esteira de um movimento ousado de inovação no mercado financeiro que acontece no mundo todo e com potencial de revolucionar e impulsionar de vez a economia digital, mitigando a necessidade de intermediários, além de possibilitar pagamentos instantâneos (como o PIX\footnote{Hoje o PIX tem mais de 122 milhões de usuários ativos, ou 57\% da população brasileira. Graças a essa inovação, 40\% dos usuários fizeram sua primeira transferência eletrônica, o que denota o enorme potencial da tecnologia para inclusão financeira. {\bf \textcolor{myGreen03}{\citep{valente_01}}}. }), maior velocidade e menor custo de transação, alta escalabilidade e maior inclusão financeira\footnote{Esse é um ponto importante para o Brasil, com elevado número de ``desbancarizados funcionais''. Apesar do Brasil ter 189 milhões de cidadãos com relação com o SFN (Sistema Financeiro Nacional), uma grande parcela da população não possui acesso a crédito ou grandes movimentações financeiras em investimentos e similares. Veja {\bf \textcolor{myGreen03}{\citep{bcb_01}}} para essas estatísticas.}. Redução de lavagem de dinheiro e corrupção, de modo geral, também são potencialidades esperadas se as CBDCs forem adotadas. Onze países já têm a versão digital de suas moedas fiduciárias, e outros estão em alguma fase de estudos ou desenvolvimento.  Ver, por exemplo, um infográfico em \href{https://www.atlanticcouncil.org/cbdctracker/}{{\bf \textcolor{myGreen03}{Atlantic Council CBDC tracker}}}.

A implantação das CBDCs, no entanto, não é livre de desafios tecnológicos. As principais preocupações tecnológicas são relativas a ciberataques e outras brechas de segurança, como a privacidade dos dados de usuário ou entidade que transaciona esses ativos. A ameaça potencial à privacidade reside no fato de os bancos centrais poderiam, em princípio, monitorar e rastrear todas as transações financeiras de um indivíduo ou instituição; já a questão dos ciberataques, a preocupação é que um ataque bem-sucedido à CBDC poderia resultar no roubo de vastas quantidades de dinheiro fiduciário, causando uma grande disrupção econômica.


Diversas empresas nacionais e internacionais, reconhecidas por sua {\it expertise} {\bf \textcolor{myGreen03}{\citep{mancini_01, gartner_02}}}, têm se debruçado sobre esses problemas e iniciaram uma série de experimentos e programas em pesquisa aplicada em computação confidencial para resolvê-los. {\bf \textcolor{myGreen03}{\citep[Ver, por exemplo,][]{microsoft_02,microsoft_04,bbchainResearch}}}.


Ao longo deste artigo, falaremos sobre novas abordagens tecnológicas promissoras para endereçar esses problemas, mergulhando no conceito de computação confidencial dentro do {\it framework} de redes distribuídas e do {\it Confidential Consortium Framework}, da Microsoft.

Em especial, falaremos sobre um desses experimentos -- feito com um parceiro confidencial -- no qual foi criada uma aplicação em Hyperledger Besu para tratar de compra e venda de ativos digitais usando como meio de pagamento o análogo a moedas digitais fiduciárias, em outras palavras, CBDCs {\bf \textcolor{myGreen03}{\citep{bbchainResearch}}}.

\section{Privacidade, proteção e a computação confidencial}

As CBDCs geralmente utilizam Blockchain ou DLTs\footnote{Na realidade, Blockchain é um tipo de DLT. Veja {\bf \textcolor{myGreen03}{\cite{anwar_01}}} para uma explicação mais detalhadas sobre os tipos e classes de DLTs e suas semelhanças e diferenças.}, tecnologias descentralizadas e seguras para transacionar, registrar e verificar ativos digitais. Essas tecnologias criam um registro digital compartilhado de transações, mantido e validado por uma rede distribuída de nós. A segurança é garantida através de técnicas criptográficas, como assinaturas digitais e funções {\it hash}.

Para proteger a confidencialidade e privacidade, são usadas, por exemplo, redes privadas com membros autorizados, criptografia de dados e pseudoanonimização das identidades dos participantes. Entretanto, é preciso considerar o equilíbrio entre privacidade, segurança e usabilidade, já que nenhum sistema é 100\% seguro.

A encriptação usual de dados sensíveis os protege apenas até o momento em que precisam ser processados, quando são decriptados e expostos a ameaças. Por essa razão, entre outras, moedas digitais, como CBDCs e {\it stablecoins}, demandam cuidados adicionais em segurança de dados, principalmente com a adoção em larga escala por instituições financeiras.

A computação confidencial é um paradigma que protege dados sensíveis enquanto estão sendo processados em ambientes não confiáveis, como nuvens públicas ou centros de dados compartilhados {\bf \textcolor{myGreen03}{\citep[ver, por exemplo,][]{costan_01,confidential_computing_consortium}}}. A BBChain é uma das empresas que está explorando esse paradigma computacional para aprimorar a segurança de suas soluções enquanto sugere que instituições financeiras utilizem esse {\it framework} como {\it benchmark} para ativos digitais.

\subsection{O \textbf{\textit{Confidential Consortium Framework}} (CCF)}

\begin{figure}[!h]
    \centering
    \includegraphics[width=9.0cm, height=5.625cm]{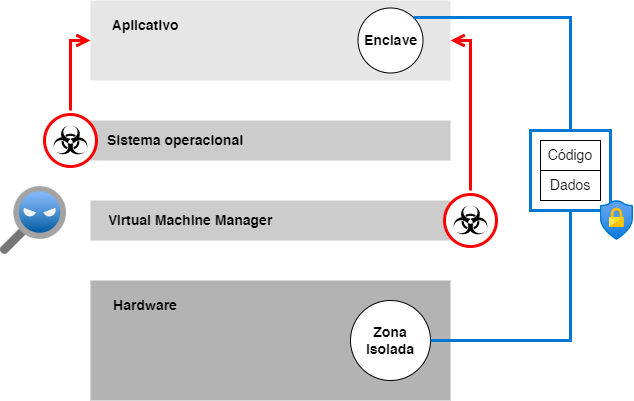}
    \caption{Abstração da proteção fornecida quando se cria um Enclave. O processador aloca uma porção da memória, isolada de qualquer outro processo, dedicada à aplicação e seus dados. Dados no Enclave são encriptados {\bf \textcolor{myGreen03}{\citep{bbchainResearch}}}.}
    \label{fig:enclave}
\end{figure}

A computação confidencial utiliza {\it Trusted Execution Environments} (TEEs) como Intel SGX {\bf \textcolor{myGreen03}{\citep{intel_03}}} ou AMD SEV {\bf \textcolor{myGreen03}{\citep{amd_01}}}, criando Enclaves onde dados sensíveis são processados de forma segura e isolada. A {\bf \textcolor{myGreen03}{Figura \ref{fig:enclave}}} ilustra o conceito de Enclave.

Enclaves são regiões específicas de memória no computador onde dados sensíveis podem ser processados, totalmente protegidos de ameaças externas como {\it malwares} ou ataques cibernéticos. Especificamente, o processador aloca uma parte da memória, isolada de qualquer outro processo e dedicada a um aplicativo e seus dados, no momento da inicialização do dispositivo. Os dados no Enclave são criptografados para evitar espionagem ou exposição, mesmo quando processados por um aplicativo. Assim, mesmo que o sistema operacional ou hipervisor sejam comprometidos, o aplicativo e os dados em execução no Enclave estarão seguros.

Por essa característica fundamental, aplicações para computação confidencial incluem saúde, finanças, supply chain e governo. Esse paradigma é útil em ambientes distribuídos, como Blockchain/DLTs, onde nem todos os participantes confiam uns nos outros.

A BBChain é uma das empresas de soluções em Blockchain/DLT, que vem explorando a computação confidencial em suas soluções, desenvolvendo várias provas de conceito (PoCs) e experimentos que encontram os requisitos funcionais e não funcionais para a implantação de um sistema de CBDC como o que se propõe a ser o Real Digital. Assim, estudos têm sido realizados utilizando o {\it Confidential Consortium Framework} (CCF) da Microsoft {\bf \textcolor{myGreen03}{\citep{microsoft_03, microsoft_04}}}.

O CCF é um {\it framework open-source} para serviços {\it stateful} de alta disponibilidade, combinando a facilidade de uso e desempenho da computação centralizada com a confiança descentralizada. Ele permite a execução auditável em dados confidenciais sem a necessidade de confiança mútua entre os participantes.

\begin{figure}[!h]
    \centering
    \includegraphics[width=12.0cm, height=6.0cm]{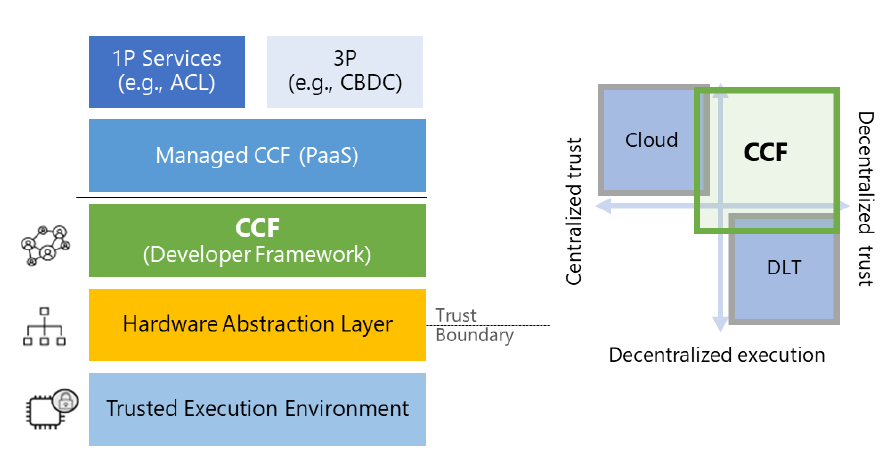}
    \caption{À esquerda, CCF na pilha de execução; à direita o CCF relativamente a outras DLTs e sistemas centralizados {\bf \textcolor{myGreen03}{\cite{microsoft_02}}}.}
    \label{fig:CCFeDLTs}
\end{figure}

A {\bf \textcolor{myGreen03}{Figura \ref{fig:CCFeDLTs}}} ilustra a combinação de características de redes centralizadas e descentralizadas no CCF, enquanto a {\bf \textcolor{myGreen03}{Tabela \ref{tabelaComparativa_BDT_DLT_CCF}}} compara três possibilidades de implementação.

\setlength{\arrayrulewidth}{0.75mm}
\setlength{\tabcolsep}{15pt}
\renewcommand{\arraystretch}{1.5}
\arrayrulecolor[HTML]{DB5800}

\begin{table}[!h]
\scriptsize 
\centering
\renewcommand{\arraystretch}{1.5}
\sffamily
\begin{NiceTabular}{ |m{4.2cm}|m{4.2cm}|m{4.2cm}| }
\CodeBefore
    \columncolor{myBlue05!100}{3}
    \rowcolor{myBlue04!100}{1}
    \rowcolor{myBrilliantGreenW!100}{2,4,6,8,10,12}    
\Body  

    \hline
    \Block{1-1}{ {\bf \textcolor{white}{Base de dados tradicional}} } & \Block{1-1}{ {\bf \textcolor{white}{DLT permissionada}}  } & 
    \Block{1-1}{ {\bf \textcolor{white}{Microsoft CCF}} }  \\

    \hline
    \Block{1-3}{ {\bf \textcolor{black}{Confidencialidade}}} \\ 

    \hline 
    \Block[l]{1-1}{ $\bullet$ O controle tradicional de acesso baseado em roles é usado para dados privados. \\ $\bullet$ Os operadores de nó precisam ser confiáveis. } &
    \Block[l]{1-1}{ $\bullet$ Todos os nós executores veem todas as transações. \\ $\bullet$ Criptografia cara é necessária para ocultar dados confidenciais. \\ $\bullet$ Muitas vezes, os operadores de nós precisam ser confiáveis para privacidade. } & 
    \Block[l]{1-1}{ $\bullet$ Controle programável sobre o que é privado no ledger. \\ $\bullet$ Apenas o órgão de governança precisa ser confiável para proteger a privacidade. \\ $\bullet$ Os operadores de nó não obtêm privilégios de leitura por padrão. } \\

    \hline
    \Block{1-3}{ {\bf \textcolor{black}{Descentralização (cibersegurança)}}} \\

    \hline 
    \Block[l]{1-1}{ $\bullet$ A governança centralizada dá imenso poder aos operadores de nós. \\ $\bullet$ Muitas vezes, a força está no elo mais fraco. Comprometer um nó pode ser suficiente. \\ $\bullet$ A operação do nó deve ser um alto privilégio. } & 
    \Block[l]{1-1}{ $\bullet$ Por meio da replicação da execução de transações. \\ $\bullet$ Cada participante deve operar um nó de execução. \\ $\bullet$ Suporta até 1/3 dos nós corrompidos. } & 
    \Block[l]{1-1}{ $\bullet$ Por meio da governança, a autoridade controla o código autorizado, os aplicativos e os operadores de nós. \\ $\bullet$ Recibos verificáveis identificam agentes mal-intencionados. \\ $\bullet$ Execute a rede em várias plataformas TEE para limitar a confiança de HW. \\ $\bullet$ A operação do nó é uma redundância simples sem privilégios. }  \\

    \hline
    \Block{1-3}{ {\bf \textcolor{black}{Disponibilidade (resiliência)}}} \\

    \hline
    \Block[l]{1-1}{ $\bullet$ Requer um hot standby e sincronização de transações em muitos casos. \\ $\bullet$ Sem cenários de falha bizantina. } & 
    \Block[l]{1-1}{ $\bullet$ Todos os nós parcialmente confiáveis para confidencialidade e integridade. \\ $\bullet$ Difícil de ser recuperado da falha bizantina. } & 
    \Block[l]{1-1}{ $\bullet$ Operadores de nó não confiáveis para confidencialidade/integridade. \\ $\bullet$ Recuperação de desastres, contando com HSM ou compartilhamentos de membros. }  \\

    \hline
    \Block{1-3}{ {\bf \textcolor{black}{Performance}}} \\    

    \hline
    \Block[l]{1-1}{ $\bullet$ Sem limite de escalabilidade em transações ou dados de transação. \\ $\bullet$ As transações são executadas apenas uma vez. \\ $\bullet$ A taxa de transferência e a latência são excelentes com réplicas e partições. \\ $\bullet$ Até 5 Lac TPS existem. } & 
    \Block[l]{1-1}{ $\bullet$ A taxa de transferência e a latência são mal dimensionadas com número de participantes. \\ $\bullet$ Todos os nós executam todas as transações. \\ $\bullet$ 20.000 TPS sem proteção de privacidade. } & 
    \Block[l]{1-1}{ $\bullet$ Nenhum limite de escalabilidade no consórcio de governança. $\sim$ 60.000 TPS foram alcançadas. \\ $\bullet$ A taxa de transferência e a latência são dimensionadas perfeitamente com a contagem de réplicas. As transações são executadas apenas uma vez. \\ $\bullet$ As partições requerem uma lógica especial que incorre em compensações de desempenho. }  \\

    \hline
    \Block{1-3}{ {\bf \textcolor{black}{Auditabilidade}}} \\    

    \hline
    \Block[l]{1-1}{ $\bullet$ É necessário criar um mecanismo especial de verificação de auditoria. Não há garantias à prova de violação. \\ $\bullet$ Mesmo assim, os operadores de nó precisam ser confiáveis. } & 
    \Block[l]{1-1}{ $\bullet$ Todas as transações são verificáveis no ledger; inviolável se houver consenso. \\ $\bullet$ Necessidade de verificar a consistência de 2/3 dos nós para provar que uma transação foi executada. } & 
    \Block[l]{1-1}{ $\bullet$ Todas as transações, incluindo a governança, verificáveis no ledger. \\ $\bullet$ Recibos off-line universalmente verificáveis com garantias à prova de violação. } \\

    \hline
    \Block{1-3}{ {\bf \textcolor{black}{Flexibilidade}}} \\    

    \hline
    \Block[l]{1-1}{ $\bullet$ Mais adequado para o modelo de conta. UTXO requer representação complexa e verificações de proveniência tornam as operações muito lentas. } & 
    \Block[l]{1-1}{ $\bullet$ Mais adequado para o modelo UTXO. } & 
    \Block[l]{1-1}{ $\bullet$ Suporta UTXO e modo de conta. } \\
    

    \hline
\end{NiceTabular}

\caption{Comparativo entre bases de dados tradicionais, DLTs permissionadas e CCF {\bf \textcolor{myGreen03}{\citep{microsoft_02}}}.}
\label{tabelaComparativa_BDT_DLT_CCF}

\end{table}

O CCF possui características similares às DLTs, como governança por consórcio de Membros regidos por uma Constituição e uma rede formada por Nós. Os Nós devem executar sobre um TEE, garantindo descentralização e segurança. O protocolo de consenso adotado pelo CCF é o Crash Fault Tolerance (CFT), baseado no Raft.

Mais especificamente:

\begin{itemize}
    \item {\bf Governança:} a rede do CCF é governada por um consórcio de Membros regidos por uma Constituição. A Constituição é como um smart contract universal composto por um conjunto mínimo de JavaScripts que formam um módulo que, por sua vez, é registrado no próprio Ledger. É a constituição que dita as regras da rede com relação às transações, smart contracts etc. Os Membros, de forma geral, governam a rede e suas identidades públicas são registradas no framework. Membros diferem fundamentalmente dos Operadores, que tratam da parte operacional da rede, como adicionar ou remover os Nós, enquanto suas identidades não são registradas no CCF.
    \item {\bf Nós:} os Nós formam a rede CCF. O entendimento é o mesmo que se tem quando falamos de uma Blockchain ou DLT. São pessoas físicas ou jurídicas com interesse legítimo de transacionar ativos dentro da rede. A especialidade desses Nós, no entanto, é que cada um deles deve rodar obrigatoriamente sobre um TEE, ou seja, em um Enclave. A rede é, portanto, descentralizada. Os Nós rodam todos os mesmos aplicativos escritos em JavaScript ou C++.
    \item {\bf Protocolo de Consenso:} Blockchains e DLTs suportam muitos tipos de protocolos de consenso, como Proof-of-Work, Proof-of-Stake, Proof-of-Authority e muitos outros. O CCF, como uma rede descentralizada que precisa garantir confiança, suporta o Crash Fault Tolerance (CFT). O CFT é baseado no Raft, mas com algumas diferenças-chave. É possível implementar também o protocolo BFT, Bizantine Fault Tolerance.
\end{itemize}

A inicialização de uma rede CCF envolve a geração de identidades, inicialização do Nó inicial e adicionais, e a proposição de abertura da rede. A governança da rede é estabelecida por uma constituição, formada por scripts em JavaScript que exportam funções de validação, resolução e aplicação.

\begin{figure}[!h]
    \centering
    \includegraphics[width=16.0cm, height=8.5cm]{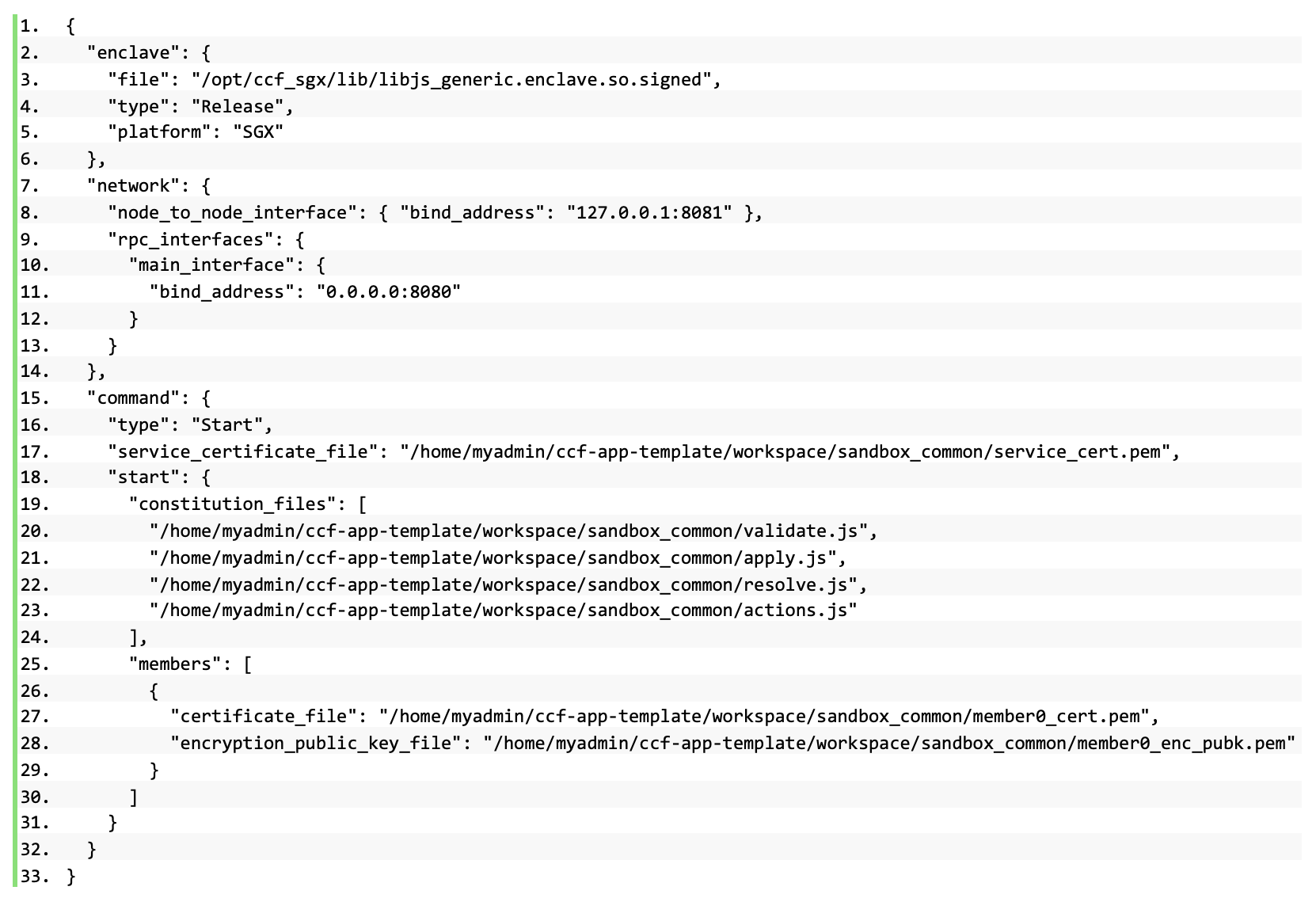}
    \caption{Demonstração do arquivo de configuração com o comando ``Start'' {\bf \textcolor{myGreen03}{\citep{bbchainResearch}}}.}
    \label{fig:demoComandoStart}
\end{figure}

\begin{figure}[!h]
    \centering
    \includegraphics[width=16.0cm, height=8.5cm]{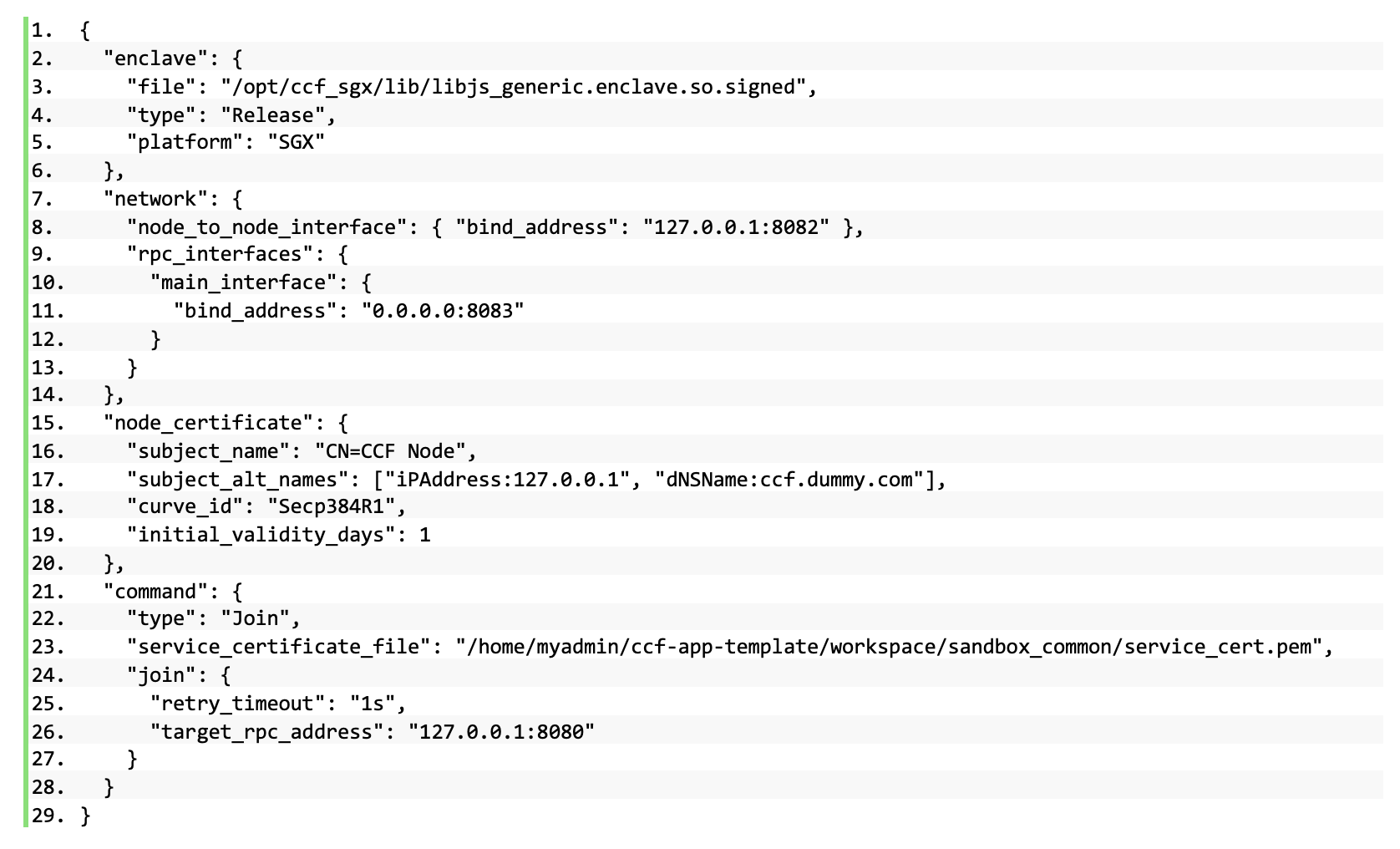}
    \caption{Demonstração do arquivo de configuração com o comando ``Join'' {\bf \textcolor{myGreen03}{\citep{bbchainResearch}}}.}
    \label{fig:demoComandoJoin}
\end{figure}

As {\bf \textcolor{myGreen03}{Figuras \ref{fig:demoComandoStart} e \ref{fig:demoComandoJoin}}} exemplificam a inicialização e inclusão de Nós na rede CCF. Com isso, abordamos os conceitos fundamentais da computação confidencial no CCF, que se mostra versátil para transações de ativos financeiros.

\section{Pesquisa e desenvolvimento em computação confidencial}

\begin{figure}[!h]
    \centering
    \includegraphics[width=9.0cm, height=9.0cm]{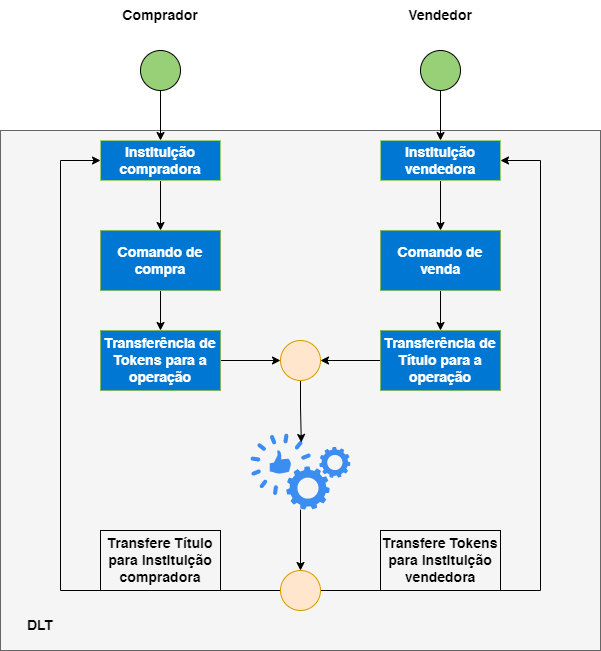}
    \caption{Aplicativo de compra e venda de ativos simulando o uso de moedas digitais {\bf \textcolor{myGreen03}{\citep{bbchainResearch}}}.}
    \label{fig:arquiteturaPoCaltoNivel.drawio}
\end{figure}

A {\bf \textcolor{myGreen03}{Figura \ref{fig:arquiteturaPoCaltoNivel.drawio}}} revela a arquitetura, em alto nível, do experimento discutido neste artigo {\bf \textcolor{myGreen03}{\citep{bbchainResearch}}}. O experimento versa sobre a emissão de títulos para compra e venda no atacado entre instituições financeiras.

Esse experimento usa um modelo de Hyperledger Besu sobre o qual roda um {\it smart contract}. Esse desenvolvimento provará que essa DLT, e o sistema criado, será apropriado para esse tipo de transação de ativos, que envolveria, por exemplo, títulos públicos e {\it stablecoins}/CBDCs.

A computação confidencial é uma linha de pesquisa concomitante e paralelo ao experimento com o Hyperledger Besu na qual se usa o {\it Confidential Consortium Framework} -- CCF. 

\begin{figure}[!h]
    \centering
    \includegraphics[width=16.0cm, height=6.0cm]{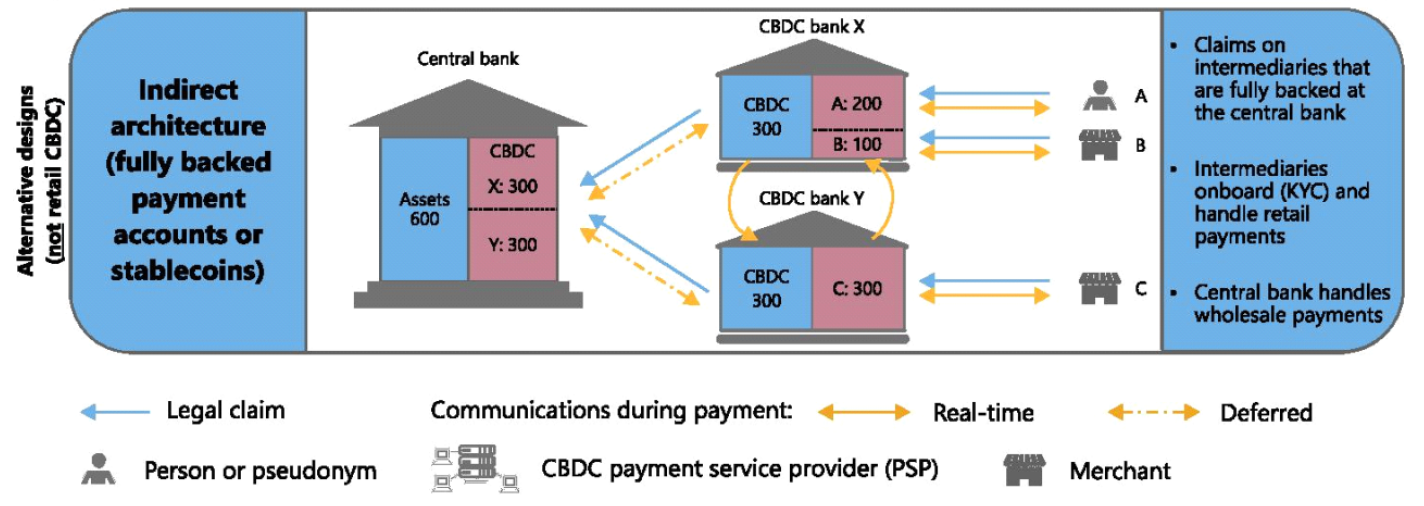}
    \caption{Na arquitetura indireta, um CBDC é emitido e resgatado apenas pelo banco central, mas isso é feito indiretamente aos intermediários. Os intermediários, por sua vez, emitem uma reclamação aos consumidores. O intermediário é obrigado a apoiar totalmente cada reivindicação com uma posse da CBDC no banco central. O banco central opera apenas o sistema de pagamentos por atacado {\bf \textcolor{myGreen03}{\citep{auer_01}}}.}
    \label{fig:modeloIndiretoCBDC}
\end{figure}

No estágio atual da pesquisa, o intuito é desenvolver um novo experimento objetivando o modelo indireto de CBDCs no atacado, sendo este modelo arquitetural um dos quatro modelos visionado pelo BIS {\bf \textcolor{myGreen03}{\citep{auer_01}}}, como mostrado na {\bf \textcolor{myGreen03}{Figura \ref{fig:modeloIndiretoCBDC}}}, tirada do referido artigo.

Assim, a ideia é usar o CCF para mimetizar essa arquitetura tendo em vista os movimentos do Banco Central do Brasil em direção a uma CBDC e de outras instituições com relação a {\it stablecoins}.

Resultante dessa pesquisa, esperamos metrificar o desempenho das abordagens e inserir, no {\it roadmap} tecnológico, o CCF como opção ao mercado {\bf \textcolor{myGreen03}{\citep{bbchainResearch}}}.

\subsection{Metodologia}

~

{\bf Definição do problema:} O uso de DLTs para transacionar dados financeiros requer cuidados adicionais para que privacidade, segurança transacional, confidencialidade e performance sejam atingidos em níveis satisfatórios, especificamente, retirando a possibilidade de uma autoridade monetária, e.g. o Banco Central, rastrear movimentações financeiras de um indivíduo identificado. 

{\bf Hipótese:} Implementar computação confidencial sobre as DLTs resolveria o problema de privacidade e de segurança, pois os dados seriam encriptados mesmo enquanto em processamento nos TEEs.

{\bf Design da pesquisa propriamente dito:} 

\begin{itemize}
    \item Acionar o suporte aos TEEs no experimento em Hyperledger Besu, implementando a computação confidencial diretamente nessa DLT;
    \item Modificação do experimento em Hyperledger Besu para incorporar o CCF para o caso de computação confidencial;
    \subitem Refatorar o {\it smart contract} para garantir sua compatibilidade com o CCF (arquitetura e funcionalidade).
    \item Uso de métricas de performance (KPIs) para comparação entre os dois cenários:
    \subitem Throughput de transação;
    \subitem Latência;
    \subitem Consumo de recursos computacionais;
    \subitem Escalabilidade.
    \item Uso de métricas de segurança:
    \subitem Resistência a ataques;
    \subitem Garantia de privacidade;
    \subitem Confidencialidade de dados durante a fase de processamento.
    \item Comparação entre as soluções:
    \subitem Solução em Hyperledger Besu sem computação confidencial;
    \subitem Solução em Hyperledger Besu com computação confidencial;
    \subitem Solução integrada ao CCF.
    \item Análise de indicadores.
\end{itemize}

\subsection{O progresso \textbf{\textit{so far}}}

O ambiente Hyperledger Besu está funcional e operante, abrigando a aplicação para compra e venda de ativos via o análogo de CBDCs.

Está sendo implementado no \href{https://www.blockchainlab.network}{{\bf \textcolor{myGreen03}{BlockchainLab}}} um ambiente com computação confidencial usando CCF como prova de conceito, inicialmente através de aplicações no estilo {\it Hello World}, validando todos os processos de governança e auditoria {\bf \textcolor{myGreen03}{\citep{bbchainResearch}}}.

\begin{figure}[!h]
    \centering
    \includegraphics[width=16.0cm, height=6.0cm]{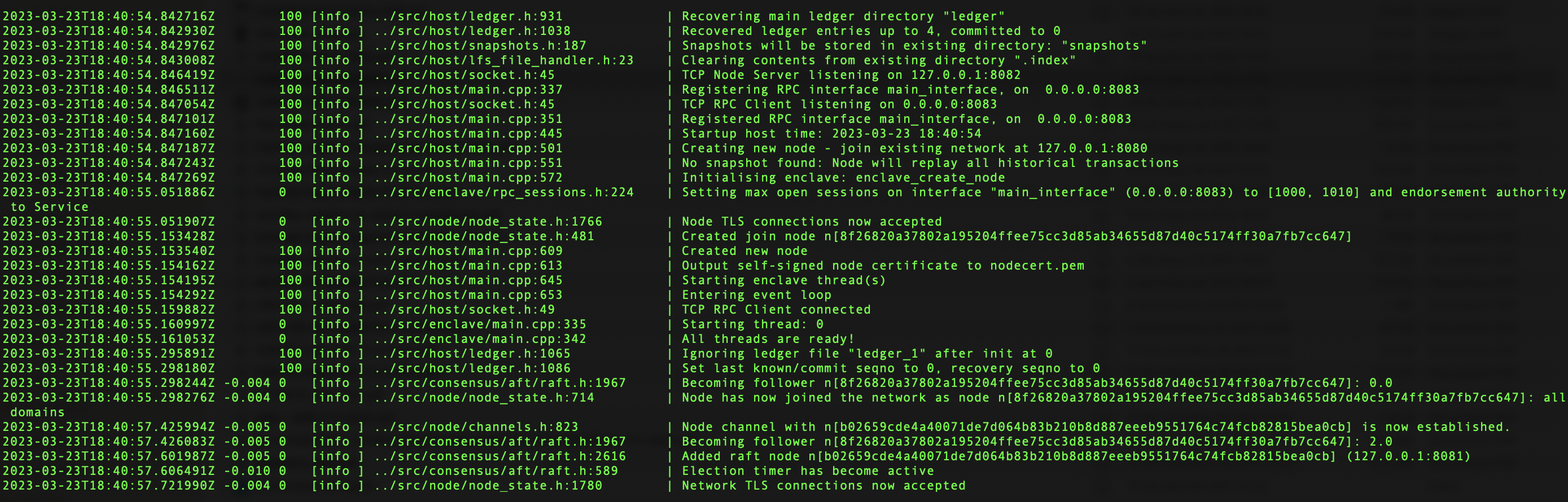}
    \caption{Nó entrando na rede como {\it leader} {\bf \textcolor{myGreen03}{\citep{bbchainResearch}}}.}
    \label{fig:noLider}
\end{figure}

\begin{figure}[!h]
    \centering
    \includegraphics[width=16.0cm, height=6.0cm]{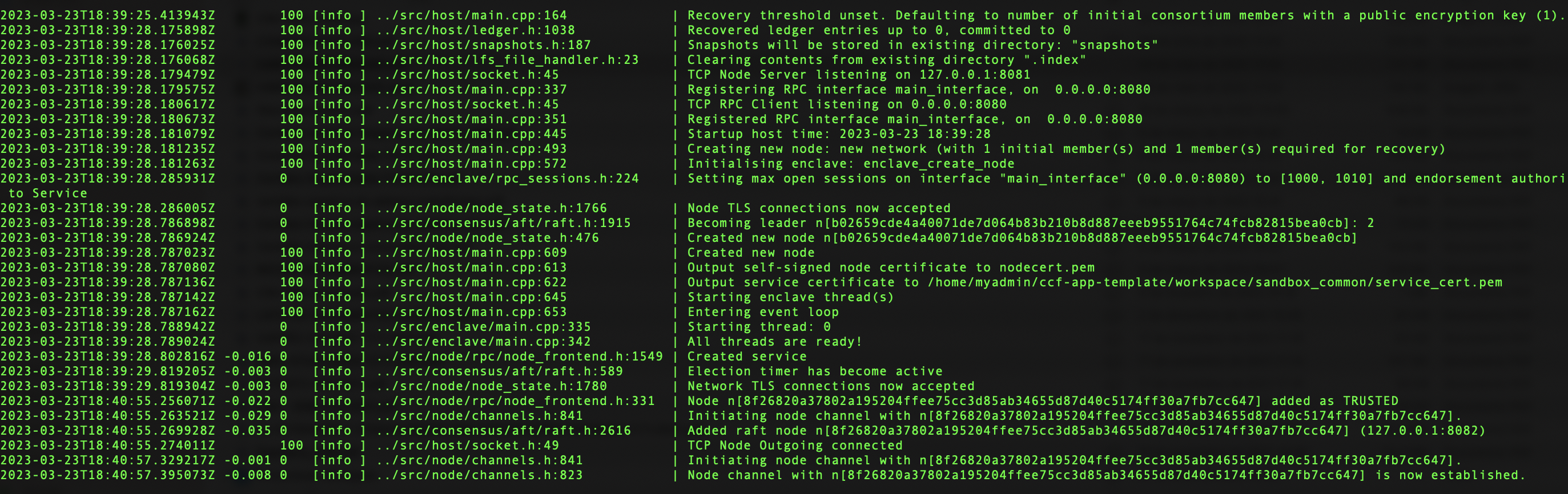}
    \caption{Nó entrando na rede como {\it follower} {\bf \textcolor{myGreen03}{\citep{bbchainResearch}}}.}
    \label{fig:noSeguidor}
\end{figure}

As {\bf \textcolor{myGreen03}{Figuras \ref{fig:noLider} e \ref{fig:noSeguidor}}} mostram as subidas dos Nós no sistema do \href{https://www.blockchainlab.network}{{\bf \textcolor{myGreen03}{BlockchainLab}}}.

O próximo objetivo da equipe de pesquisa é a construção do aplicativo de compra e venda de ativos como estudo de caso para compararmos com a aplicação operante no Hyperledger Besu. Por meio desse estudo, teremos base para avaliar os pontos positivos e negativos de cada ambiente em situação análoga, a fim de ajudar todos os interessados na escolha do {\it framework} ideal para sua realidade.

\section{Conclusão e perspectiva}

Computação confidencial e os {\it frameworks} que a implementam representam um grande avanço em como dados e códigos podem ser protegidos e compartilhados. Suas aplicações potenciais perpassam diversos segmentos industriais, como o mercado financeiro, mercado de saúde, e governos. 

Especialmente para o segmento de finanças e economia digital, de modo mais abrangente, a computação confidencial aumentará a segurança e a privacidade de CBDCs ao permitir a execução segura de transações e assegurando que dados pessoais e confidenciais jamais sejam acessados por qualquer ator não expressamente autorizado, nem mesmo os grandes provedores e custodiantes desses dados, como Microsoft, Google, AWS e Meta.

Dois outros aspectos da computação confidencial que são importantes no contexto das CBDCs são a questão da interoperabilidade com outros sistemas e plataformas – outros meios de pagamento, como PIX, por exemplo – permitindo até mesmo transações internacionais mais ágeis e seguras; o segundo aspecto está relacionado à {\it compliance} regulatória. Este ponto é de grande interesse para governos, dado que o sistema seria inerentemente aderente às regras antilavagem de dinheiro (AML) e de ``conhecer seu cliente'' (KYC).

O Banco Central do Brasil, como já mencionamos, iniciou a fase de pilotos para a implantação do Real Digital em Hyperledger Besu {\bf \textcolor{myGreen03}{\citep{bcb_03}}}.

Hyperledger Besu foi escolhida como tecnologia subjacente segundo critérios técnicos, mas não demanda a computação confidencial para salvaguardar as informações e dados das transações. Essa segurança deverá usar outras técnicas de cibersegurança.

Na arquitetura inicial do Banco Central do Brasil, ativos digitais seriam negociado no atacado, entre instituições financeiras como bancos comerciais e outras, utilizando o CBDC do Real Digital.

Assim, conforme a economia digital se expande, a sociedade enfrentará uma decisão crucial em relação ao futuro do dinheiro digital: ter de escolher entre formas soberanas e não soberanas de moeda. A moeda soberana engloba o dinheiro dos Bancos Centrais (CBDCs), o dinheiro de bancos comerciais e o dinheiro eletrônico emitido por instituições, financeiras ou não, regulamentadas; todas essas formas são classificadas como passivos regulamentados {\bf \textcolor{myGreen03}{\citep{rln_01}}}.

Ao utilizar a tecnologia de registros compartilhados, esses instrumentos legais podem ser representados sem alterar significativamente as regras e regulamentações existentes, abordando potencialmente um grande desafio do setor: a ausência de um sistema de liquidação financeira global, multimoeda e multiativos.

Essa base abre caminho para uma nova proposta de Infraestrutura do Mercado Financeiro (FMI) chamada Rede de Responsabilidade Regulamentada (RLN, na sigla em inglês). A RLN operaria um registro compartilhado que registra, transfere e liquida passivos regulamentados de bancos centrais, bancos comerciais e instituições regulamentadas. Por exemplo, à medida que as {\it stablecoins} se tornam reconhecidas como uma forma de passivo regulamentado em algumas jurisdições, elas também podem ser integradas à rede, permitindo a interoperabilidade com outras formas de dinheiro tokenizado, especialmente, as CBDCs.

O principal objetivo da RLN é estabelecer uma infraestrutura de registro compartilhado para o sistema de moeda soberana que seja continuamente operacional, programável e multiativos. Essa rede facilitaria a finalização ``on-chain'' de liquidação entre instituições participantes em moedas soberanas, mantendo-se em conformidade com todas as regras e regulamentações existentes {\bf \textcolor{myGreen03}{\citep{rln_01}}}.

É nesse contexto que o CCF pode encontrar sua maior aplicabilidade. Uma RLN, em suma, é um conceito no qual uma rede de participantes adere a regras e regulamentações específicas impostas por um órgão regulador. Essas redes geralmente lidam com dados sensíveis e exigem comunicação segura entre os participantes, além da capacidade de rastrear e auditar transações.

Assim, visionamos que o CCF poderia ajudar a evoluir o conceito de RLN das seguintes maneiras:

\begin{itemize}
    \item Maior Privacidade e Segurança: os TEEs seriam benéficos em RLNs, onde privacidade e segurança são fundamentais.
    \item Confiança e Conformidade Aprimoradas: o CCF possibilita a criação de aplicativos descentralizados com alto grau de confiança, já que os participantes podem verificar a correta execução do código dentro dos TEEs. 
    \item Escalabilidade e Desempenho: O CCF foi projetado para oferecer alta taxa de transferência e baixa latência, tornando-o adequado para RLNs em larga escala. Além disso, a capacidade da plataforma de lidar com um grande volume de transações sem sacrificar o desempenho pode promover a adoção de aplicativos descentralizados nas RLNs.
    \item Interoperabilidade: O CCF pode ser integrado a várias plataformas de blockchain e outras tecnologias de registro distribuído, permitindo a interação perfeita entre diferentes sistemas. Essa interoperabilidade facilitará a colaboração entre RLNs, reguladores e outros interessados no ecossistema.
    \item Auditoria e Monitoramento: os recursos integrados do CCF, como logs à prova de violação, permitem um processo de auditoria eficiente e transparente, de modo que as RLNs cumpririam automaticamente os requisitos regulatórios e fornecer informações para melhoria contínua.
    \item Modelos Flexíveis de Governança: O CCF pode acomodar diferentes estruturas de governança, permitindo que as RLNs implementem modelos de governança personalizados de acordo com suas necessidades regulatórias e organizacionais.
\end{itemize}

Em especial a interoperabilidade pode ser útil para integração com outros sistemas de pagamento eletrônico como PIX no Brasil, sem falar que o Banco Central brasileiro escolheu a DLT Hyperledger Besu para os pilotos do Real Digital.
    
Em conclusão, o CCF poderia contribuir significativamente para a evolução das Redes de Responsabilidade Regulamentada, fornecendo maior privacidade, segurança e conformidade, ao mesmo tempo em que promove escalabilidade, interoperabilidade e modelos de governança flexíveis. 

Empresas brasileiras vêm investindo maciçamente na formação de talentos capazes de aprimorar as tecnologias Blockchain/DLT e implementar sistemas de grande impacto social no Brasil {\bf \textcolor{myGreen03}{\citep{bbchainResearch}}} e no mundo.


Notadamente, plataformas como o \href{https://www.blockchainlab.network}{{\bf \textcolor{myGreen03}{BlockchainLab}}}\footnote{Para mais informações sobre o \href{https://www.blockchainlab.network}{{\bf \textcolor{myGreen03}{BlockchainLab}}}, suas capacidades, inovações e processo de ideação, ver {\bf \textcolor{myGreen03}{\cite{deAvellarMGB_02}}}, em sua Monografia apresentada na Especialização em Gestão Estratégica da Inovação Tecnológica, no Departamento de Política Científica e Tecnológica da Universidade Estadual de Campinas, Unicamp.}, com toda sua versatilidade tecnológica e governança {\it as a Service}, se posicionam na fronteira da tecnologia de registros distribuídos, seguindo de perto os novos paradigmas computacionais.


\section{Agradecimentos}
Os autores agradecem à BBChain pelo incentivo e à Microsoft pela oportunidade.

\noindent Os autores também agradecem a Davi Castelo Branco Dias da Cunha, da Microsoft, pela revisão do artigo.

\section{Conflito de interesse}

Não há conflito de interesses no que se refere a essa pesquisa.

\clearpage

\bibliography{bibliografia}

\end{document}